\documentclass[a4, 11pt]{article}
\usepackage[textwidth=16cm]{geometry}
\usepackage{amssymb, amsmath}
\usepackage{bm, graphicx}
\usepackage{physics}
\usepackage{color,cite,authblk,ulem}
\usepackage{multirow}

\newcommand{\fig}[1]{Fig.~(\ref{#1})}

\begin{document}
\title{Emergence in Science} 
\author{Nitish Kumar Gupta, A. M. Jayannavar}

%\affiliation{Centre for Lasers \& %Photonics, Indian Institute of Technology %Kanpur, 208016, India}

\maketitle
\begin{abstract}
 In the scientific literature, the term emergent phenomena is invoked in the context of a collective behavior observed in a complex adaptive system that exhibits no correspondence with the behavior of the system constituents. Although this description is too generic to be termed a definition, it alludes to the fact that emergent phenomena are closely related to esoteric observations and neoteric developments in science. With these impressions, we aim to investigate a variety of observed behavior from the perspective of an emergentist. Starting with a few familiar portrayals of emergence, we devote a large body of this narrative review to explain instances of emergence in condensed matter systems, namely the phase transition phenomena, spontaneous symmetry breaking, and macroscopic quantum phenomena. At the same time, to present a broader perspective, we cross the domain boundaries to also provide a succinct description of emergent phenomena prevailing in social sciences, economics, computing environments, and in biological systems, as well.
\end{abstract}

%\linenumbers
\section*{Introduction}

Any difference between a tree and a human being is impossible to tell at the level of individual atoms. The difference comes in how those same atoms are organized, yielding bark, leaves, and roots on the one hand and eyes, hair, blood, and organs on the other. Yes, it sounds curious, but usually, we take for granted the fact that the world around us, although composed of the same limited building blocks (the 92 stable atoms), is structured in multiple ways. As we start investigating, we find that new forms of complexity, organization, and characteristics arise at every level of description. This is the phenomenon of emergence; it applies as much to natural and physical phenomena as it does to social and economic sciences. Similarly, its manifestations can be seen at every length and time scale, even in abstract spaces and in the domain of fine arts as well. For example, the impressionist art movement, started by a group of artists in the 19th century France, reflects all the characteristics of emergent behavior. The works of Paul Cezanne, Claude Monet, and Auguste Seurat are all examples of the mentioned art form. This group of artists experimented with a radically different way of painting. If you look at an impressionist painting close-up, you will see distinct short brushstrokes of different colors, lacking any coherence. Now, slowly move away from the painting and at some point, these brushstrokes merge into a definite form. Thus, a pattern emerges from a seemingly incongruent collection of elements; the pattern contains a well-recognized and quantified piece of information while the constituent elements are chaotic. Well, that is just one example; if we carefully look around us, we find such instances are quite common. Perhaps nature itself is an impressionist!

 With years of effort, in some instances, we have been able to develop a rudimental but important familiarity with the emergent properties. Particularly, detailed and systematic theoretical work has been done in physics over the years, which has enhanced our understanding of emergent phenomena. For example, the debate around the thermodynamic limit and incorporation of non-analyticities gives us some perspective about the scrambled but fertile turf that we have~\cite{batterman2011emergence}. This debate is very fundamental to our understanding of the commonplace phenomena such as liquid to solid phase transition (atoms, when cooled sufficiently, organize into a crystal, and as we will see later that it can be understood in terms of change that takes place in an ``order parameter"). However, there remains a lot to learn and comprehend about emergent phenomena, particularly in areas which are related to the origin of life. 

Moving on, let us understand the perspective in which a formal effort has been launched to study emergent phenomena. If we observe the chronology of any domain of science, one feature that stands out is that of Reductionism. In physics also, most of its history is concerned with going down to smaller and smaller scales and trying to fathom an underlying common (and unified) thread for all the observable phenomena. In his radical article of 1972, Philip W. Anderson, however, tried to look at things from an immaculate and pristine eye~\cite{anderson1972more}. What he noticed was something very apparent but at the same time uncanny as well; he recognized the fact that while the goal of reductionist efforts is to come up with a unified theory but ironically, such a construct will not facilitate the capability of recreating the universe. Hence, even though we would have understood the universe, we may still remain oblivious to the mechanism of the unfolding of most of the observed phenomena. In his own words, ``The ability to reduce everything to simple fundamental laws does not imply the ability to start from those laws and reconstruct the universe. In fact, what we think is fundamental may be an emergent reality.'' The basis of Anderson's argument came from years of observations which gave him an opinion that the laws in condensed matter systems are equally fundamental as those of high energy physics. That is why he mooted the idea that the reductionist approach to science is not suitable for describing the complexity of natural behavior that arises at every level of organization.  This accentuated the need for an alternative viewpoint in science where the effort is unvaryingly focused on the emergent phenomena rather than on conventional reductionist assertions. This was the starting point of a long-overdue endeavor, which still is a work in progress.

When we have such a colossal panorama in sight, the relevance of the subject matter is unambiguous, not to mention the richness of interactions and convergence of thoughts it brings about. However, at the same time, it also engenders difficulties in forging ``complete and comprehensive" definitions. Nevertheless, many works have contributed towards formalizing the discussions so that the trajectory becomes comprehensible. In this context, it becomes essential to put forth some minimalistic criteria for emergent phenomena, which in the words of Miller and Page~\cite{miller2009complex}, can be pinned down to two requirements : (a) Emergence is a phenomenon whereby well-formulated aggregated behavior arises from localized, individual behavior; (b) such aggregate behavior should be immune to reasonable variations in individual behavior. These requirements, although, have been put forward in the context of social behaviors, but are generic enough to stay relevant for other instances. Being minimalistic in nature, this criteria has its limitations and does not capture the most exciting aspects that are dominantly associated with `strong emergence.' For the sake of completeness, we also mention the characteristics of strong emergent phenomena: (a) Appearance of some novel properties and behavior which are contrived when the constituents interact and evolve as a whole; (b) the characteristics of this emergent whole is usually inexplicable by the fundamental theory that governs and describes the characteristics of the constituents. In such contexts, the discussion on emergent phenomena always brings in the notions of antireductionism, unpredictability, and novelty. On the other hand, adopting a utilitarian point of view, we notice that the macrostate language of emergent properties evinces out concepts that are meaningless at the microstate level, which, at times, allows us to look at the problems from a different perspective. In other words, focusing primarily on emergent properties provides obscurity with the constituent behavior and often brings in a great deal of simplicity in the description. This has been a guiding force behind the search for emergent behavior, even among those who do not consider themselves emergentists.

With this humble introduction, we proceed to three simplistic examples of emergent behavior before dwelling on  more specialized occurrences of emergence.

\section{Familiar Instances of Emergence}

As students, we have learned many concepts targeted to serve some specific purposes only. Nevertheless, oftentimes, if we step back and look at them from a broader perspective, we find that we are more familiar with the concept of emergence than we realize.

\subsection{Emergence in Geometry}

As we recall, our introduction to geometry started with `a point,' then came the description of a line, and then we moved on to a collection of lines. Such a progression looks obvious and mundane but let us look at it one more time:
The idea of angles makes no sense unless we start drawing a family of lines and the idea of volume remains irrelevant unless we start forming closed three-dimensional objects. Thus, at each level of complexity, some new traits emerge, which become the dominant characteristics of that description. At the level of three-dimensional objects, it makes no sense to talk about individual points, although the three-dimensional object is an aggregation of these points only. In formal language, this observation amounts to a statement like: at each higher stratum, some emergent properties surface which are absent at lower strata.

\begin{figure}
\begin{center}
  \includegraphics[width=0.57\textwidth, angle=00]{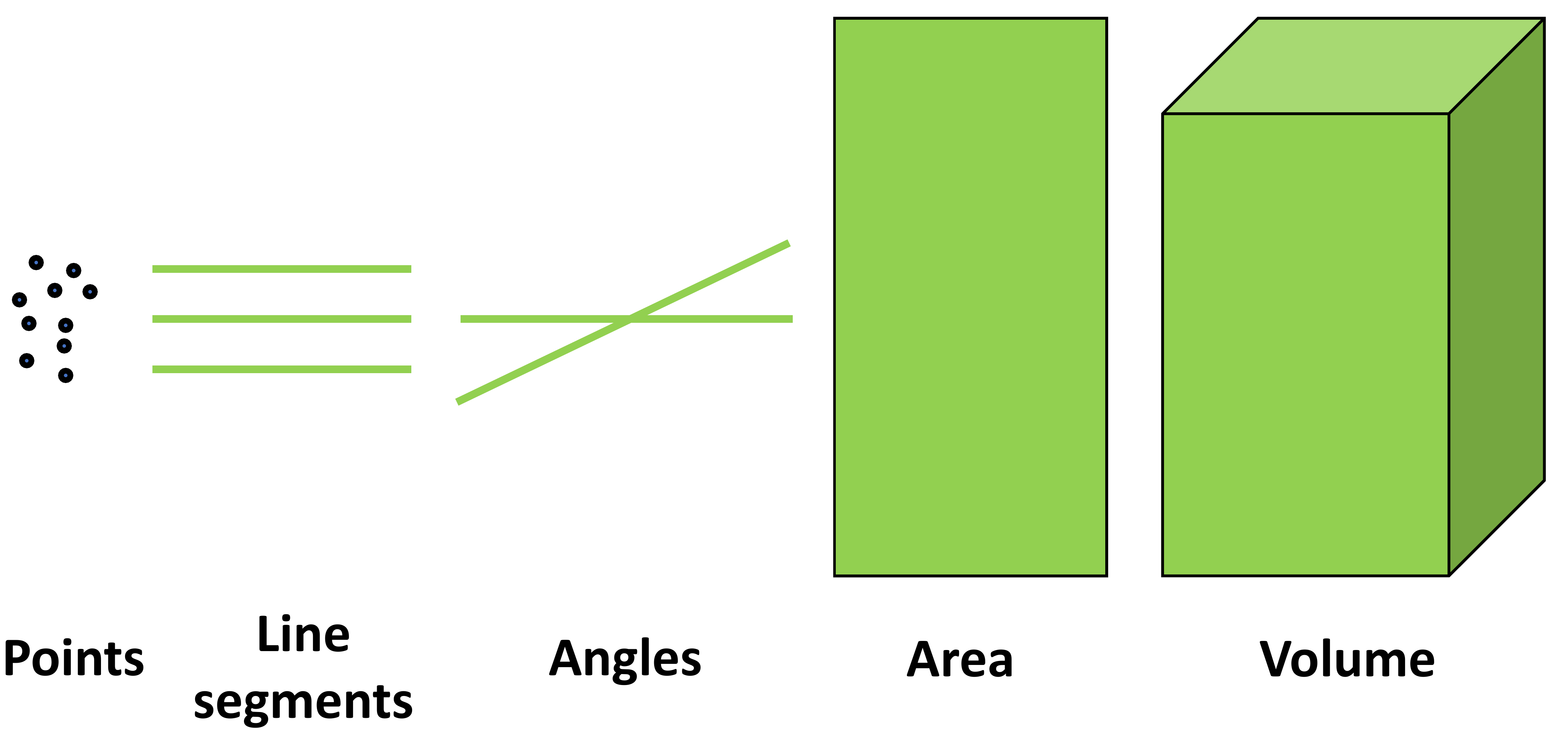}
  
  \caption{Progression of geometrical forms leads to emergence of new characteristics.}
\end{center}
\end{figure}

\subsection{The Central Limit Theorem}

The central limit theorem is a statement on the summation of probability distributions. It states that the distribution of the sum of many independent variables tends towards a normal distribution, irrespective of the distributions exhibited by the underlying variables. In other words, the sum of a large number of random variables gives rise to a Gaussian random variable. 

Consider a bag in which we place pieces of paper, each with a number written on them. There are one hundred pieces of paper, numbered from 1 to 100, in the bag. Imagine taking a piece of paper out at random, writing down the number written upon it, and then putting it back. Let us assume that we take ten such pieces of paper out, thus giving us ten numbers drawn at random, with repetition, from the range between 1 and 100. Take the sum of these numbers that are drawn and divide by 10, giving us the mean. Let us imagine that we repeat this exercise many times, each time putting back the numbers we just drew. As we might expect, each time we find the mean, we get a different number. However, if we plot the means that result, an extraordinary feature emerges. The distribution of means resembles a bell-shaped curve, peaking, as one might expect, at 50. The technical name for this bell-shaped distribution is the Gaussian or Normal distribution, and it gets generated every time we perform such an experiment, a consequence of the central limit theorem!

Now let us also dwell on some of the consequences that the central limit theorem brings about in the natural world. Phenomena in nature are invariably driven by a large number of random events, but is there a possibility of identifying an overall pattern emerging out of it?  Consider the simple case of the motion of a pollen grain on the surface of water. If we observe a pollen grain under a microscope, we may observe an irregular jittery motion of the grain. This motion is caused by the random collision of the surrounding water molecules with the pollen grains. It is the Brownian motion named after its discoverer Robert Brown (in 1827). Noting down the positions of the pollen grain as a function of time and computing the mean over a large number of observations exhibits a straightforward relation for the mean square displacement, i.e., $R^2 \propto t$. So in the case of a pollen grain, we may consider a simple random walk and average the displacement of the random walker over a large number of walks. With this exercise, we arrive at a simple ``emergent'' law which says that the mean square displacement is linearly proportional to the number of steps. Thus, a simple rule emerges out of the complex dynamics by the process of averaging.

~\\
\setlength{\fboxsep}{10pt}
\begin{center}
\fbox{
\parbox{0.8\linewidth}{
In 1908, with the help of careful observation of the motion of suspended particles in water, Jean Baptiste Perrin confirmed the prediction of Albert Einstein's quantitative theory of Brownian motion\\
\begin{align}
 \frac{1}{2}\frac{\langle x^2 \rangle}{t} = \frac{RT}{N_A} \frac{1}{6\pi a \xi}~~~~~(=D), \nonumber
\end{align}
where $D$ is the diffusion coefficient of the particle, $\langle x^2 \rangle$ the mean square displacement at time $t$, $R$ is the ideal gas constant, $T$ absolute temperature, $a$ is the radius of the Brownian particle and $\xi$ is the viscosity of the fluid. The right hand side of the above relation were obtained by the principles of kinetic theory of gases, while the left hand side quantities were obtained by observing the particle under a microscope. This experiment led to the determination of Avogadro's constant $N_A$; more importantly it finally also established the discrete nature of matter that was first proposed by John Dalton in 1808.
}
}
\end{center}

\subsection{Electron-Hole : A Well-Known ``Quasiparticle"}

Although we will be having a detailed discussion on quasiparticles when we talk of emergence in condensed matter systems; it makes sense to talk of the concept of ``electron-hole" here because of the prevailing familiarity that we have with the concept (the idea of the hole is usually first introduced as a shortcut for calculating current in a nearly filled valance band).

In the parlance of semiconductor physics, a novel quantum object, called `hole,' refers to the empty electron states in the valence band. Thus, holes are not particles in their own right and are referred to as quasiparticles (hence different from positrons). The hole based description has been developed to model the aggregate motion of electrons in the valence band of a semiconductor, and hence the properties of these quasiparticles depend on the prevailing environment, leading to their different characteristics under varying circumstances (such as their mass can be quite different from the electron mass):
\begin{equation}
    m^{*}= \hbar^2 \left[\frac{\partial^2 E}{\partial k^2}\right]^{-1}
\end{equation}

Being a representative of electron states, we hope to keep the negative sign of charge for them, but, on account of the valence band's negative curvature, the effective mass for the holes becomes negative. This leads to a quasiparticle with negative mass and negative charge, which for the sake of avoiding discomfort, is usually replaced with a particle of positive mass and positive charge. These two sign changes cancel out each other in most scenarios, avoiding the question of legitimacy. Such a description of the collective behavior of electrons offers seclusion from the intricate details of the band dynamics and provides a reasonably helpful description of the material behavior. In this sense, the emergent phenomena form a layer of abstraction that obscures us from the unsuitably complex details of the underlying phenomenon and equips us with an alternative, abstract but simpler description of the collection.

Now we move on to some detailed and domain-specific descriptions of emergent phenomena across various fields of knowledge.

\section{Emergence in Social Systems}

One of the earliest models to show emergent characteristics in social behavior was a model for urban segregation by Thomas Schelling~\cite{schelling1971dynamic}. Schelling was trying to fathom the process by which black and white communities, even if initially well-mixed across a city, might gradually become ghettoized, split up into largely homogeneous black and white localities. Schelling considered individual households as ``agents'', individual entities defined on a computer and capable of making decisions (changing their instantaneous states) depending on the input they received. He then put in a weak preference for an agent to be surrounded by agents of a similar type and allowed agents to move to new locations if their new neighborhoods accorded better with their preference. Each agent is happy only if a certain fraction of their neighbors are like them. If unhappy, they move to a vacant house. His computer simulations found that inexorably, segregation emerged, even if the tendency to prefer one type of neighbor to another was weak.

%\begin{figure}
%\begin{center}
% \includegraphics[width=0.4\textwidth, angle=-90]{Fig2_Init.eps}
%\includegraphics[width=0.4\textwidth, angle=-90]{Fig2_Fin.eps}
%  \caption{The Schelling model of social segregation: (left) initial condition, (right) emergence of segregation in the model with %threshold 
%parameter $f = 0.3$.}
%\end{center}
%\end{figure}

In later work, the social scientist Joshua M. Epstein studied a similar agent-based model for civil violence~\cite{epstein2002modeling}. He studied two variants of a simple model. In the first, a central authority seeks to suppress decentralized rebellion, while in the second scenario, a central authority seeks to suppress communal violence between two warring ethnic groups.  He took agents to be members of the general population and  considered another type of agent, the cops, to be a force of a central authority. He encoded various properties for the agents, including hardship, legitimacy, and risk aversion. Each agent performs a cost-benefit analysis, evaluating the benefits of rebellion against the costs of doing so. The results that he obtained with the first model were: Although the presence of central authority may enforce surface stability even when there is a widespread hostility against the regime, but there arise some natural `tipping points.' When situations are pushed across these tipping points, the model results in an endogenous outburst of violence. On the other hand, the outcomes of the second model were: Even with no peacekeepers, peaceful coexistence between the two ethnic groups was observed, provided there exists high `legitimacy.' With the reduction in `legitimacy' and the `force density;' however the episodes of ethnic cleansing were seen.
In yet another model, investigations are made into the question: how does modeling and understanding collective behavior help us design better ways of evacuating a stadium. The interactions of people evacuating a stadium can turn over a time of seconds into tragedy, such as in the Heysel stadium disaster. One optimal strategy for designing a stadium for effective evacuation turns out to be a counterintuitive one.  Suppose we consider evacuating a stadium with one exit. How can one increase the outflow of people? Surprisingly, the result is that a column placed just before the exit and placed slightly asymmetrically with respect to it serves to regulate the flow, leading to potentially fewer casualties and an increase in the flow. This is one more example of an emergent phenomenon.

Dwelling further on it, in the next section, we will be discussing a specific example of collective behavior and social aggregation (Flocking), which leads to the birth of swarm intelligence.

\section{Flocking and Synchronization in Natural Systems: Emergence of Swarm Intelligence}

Flocking is common in many species of organisms,  from insect swarms to schools of fish. This collective motion is only a result of interactions between members of the flock, and there is no leader who dictates the individual's behavior. For instance, in a flock of starlings, one can think of every bird's action described by simple rules. Some of these rules could be flying in the same direction as a neighboring bird, avoiding collisions with other birds, etc. Similar collective behavior is often observed in large populations of fireflies as well. The firefly is a kind of beetle that produces bioluminescence. In some parts of the world, it has been observed that a large population of fireflies often display synchronized flashing. Each firefly tries to match its flashing with neighboring fireflies. As each firefly tries to synchronize with the others, the whole population of fireflies begins to flash together. Another commonly observed phenomenon of synchronization is seen in the collective chirping of crickets. These are typical everyday examples of emergent phenomena.

 The simplest model displaying such flocking behaviour was proposed by Tamas Vicsek and collaborators in 1995~\cite{vicsek1995novel} for the flocking of birds. Each bird is described as a unit vector (an arrow) pointing along its direction of motion in the model.  At each time step, every bird is assigned to move along the average velocity of the neighboring birds within a radius $r$ with some random perturbation as added noise. They showed that when the randomness is small, the collection of vectors displays an ordered motion of the whole population. As the randomness is increased, there is a transition to disorderly motion at a critical noise level. This is an example of spontaneous symmetry breaking that can be witnessed in everyday life. ~\\
\setlength{\fboxsep}{10pt}
\begin{center}
\fbox{
  \parbox{0.8\linewidth}{
  The dynamics of the Vicsek flock is given by
  \begin{align}
     \mathbf{x}_i(t+1) = \mathbf{x}_i(t) + \mathbf{v}_i(t)\Delta t, \nonumber \\
     \theta_i(t+1) = \langle \theta_i(t) \rangle_r + \eta,\nonumber
  \end{align}
  (for all $i=1, 2,\ldots, N$), where $\mathbf{x}_i$ is the position of a particle, $\mathbf{v}_i = (\cos        \theta_i, \sin \theta_i)^T$ the velocity, $\eta$ is a random noise. The quantity $\langle \theta_i(t) \rangle_r$ is the mean of all $\theta$ within a radius $r$ of the i$^{th}$ particle. A random configuration of velocities and positions of the particles is taken as initial condition.
}
}
\end{center}

\begin{figure}
 \begin{center}
   \includegraphics[width=0.3\textwidth]{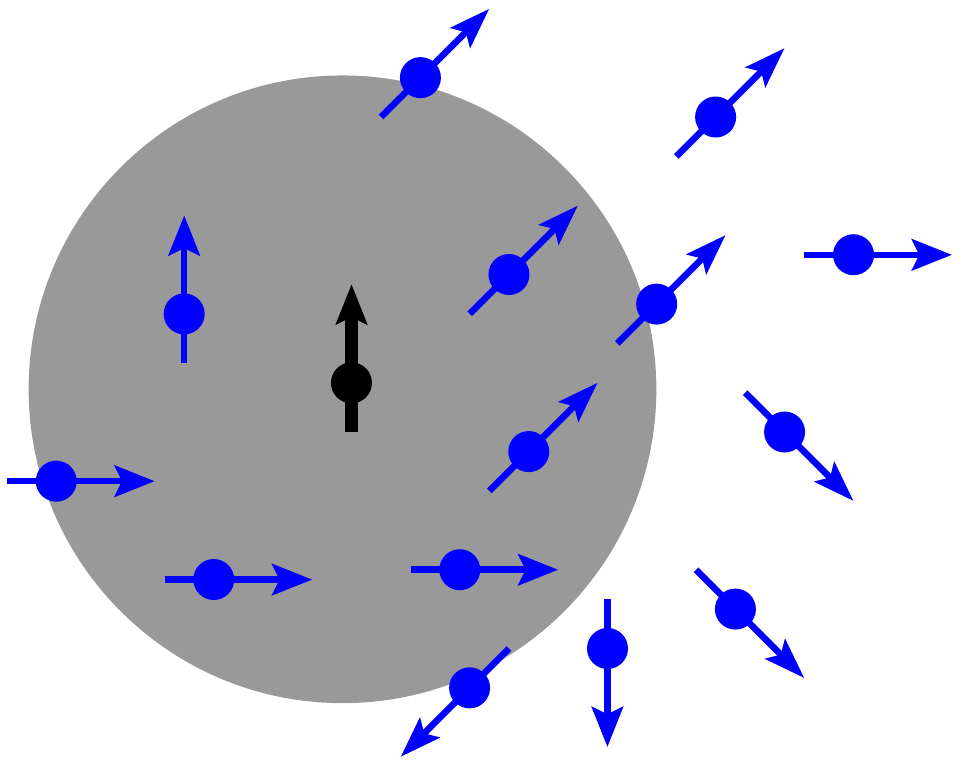}
   \caption{Formulation of Vicsek model: each point denotes a bird and the arrow denotes its direction of flight. The grey circle is the region of interaction of the bird (black) with other birds.}
 \end{center}
\end{figure}

%\begin{figure}
% \begin{center}
%   \includegraphics[width=0.7\textwidth]{flock.pdf%}
%   \caption{Simulation of Vicsek model compared to %that of a real flock (Image Source: %Wikipedia.org). Parameters: $N=1000$, $v=0.15$, %$r=1$ }
% \end{center}
%\end{figure}

In 1967 Arthur Winfree studied the collective behavior of a population of limit-cycle oscillators and proposed a mean-field model in which a spontaneous synchronization emerges, in direct analogy with phase transition (a temporal analog of phase transition)~\cite{winfree1967biological}. He had taken the natural frequencies of the coupled phase oscillators to be distributed  and assumed that each oscillator was coupled to the collective rhythm of the population. Within this framework, he discovered the collective synchronization to be a threshold phenomenon, whereupon exceeding a critical coupling strength, some oscillators exhibit spontaneous synchronization with the common frequency, thereby overcoming their inherent disorder. Yoshiki Kuramoto further refined these ideas and presented simple models of synchronization in 1975 and 1984~\cite{kuramoto1975international,kuramoto2003chemical} exhibiting dynamical phase transitions. 
%~\\
\setlength{\fboxsep}{10pt}
\begin{center}
\fbox{
  \parbox{0.8\linewidth}{
The Kuramoto model for $N$ phase oscillators is given by
\begin{align}
  \dot{\theta_i} = \omega_i + \frac{\epsilon}{N} \sum_{j=1}^N \sin(\theta_j-\theta_i),
\end{align}
 (for $i=1,\ldots,N$), where $\theta_i$ is the phase and $\omega_i$ is the natural frequency of the i$^th$ oscillator, $\epsilon$ is the strength of interaction. The complex order parameter defined as $z = N^{-1}\sum_{j=1}^N e^{i\theta_j}$ shows a transition from a completely unsynchronized ($|z| \sim 0$)oscillators at small $\epsilon$ to total synchronization ($|z| \sim 1$) at large $\epsilon$ values. For $|\omega_i - \langle \omega_i \rangle| < \epsilon |z|$ the oscillators are synchronized and the whole population is said to be `Phase Locked.'
}
}
\end{center}

\begin{figure}
\begin{center}
\includegraphics[width=0.6\textwidth, angle=0]{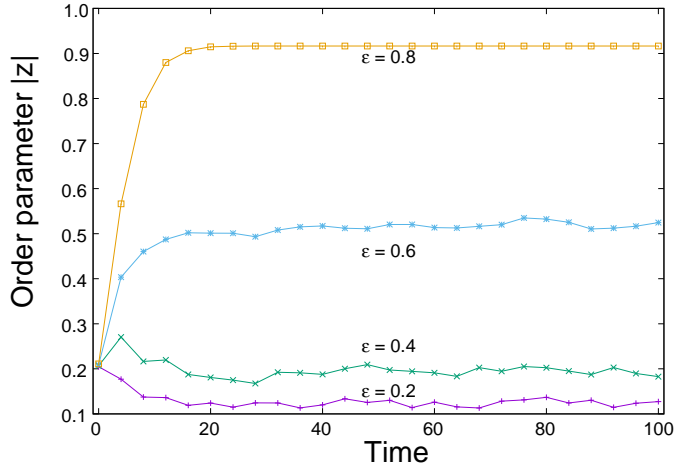}
\caption{Synchronization of Kuramoto oscillators: Order parameter characterizes various levels of synchronization for $\epsilon = 0.2, 0.4, 0.6,$ and $0.8$.}
\end{center}
\end{figure}

The  Kuramoto model consists of weakly interacting phase oscillators (with well-defined natural frequency) where each oscillator is coupled to the rest of the oscillators in the population. There are several assumptions in Kuramoto's formulation, such as the oscillators are identical, sinusoidal dependence of the interaction on the phase difference between two oscillators. Nevertheless the great utility of the model lies in the fact that this nonlinear model subscribes to analytical solutions in the limit of infinite oscillators. Over the years Kuramoto model has been proven to be very popular, and it has found widespread utility in neurosciences, chemical and biological oscillating phenomena, and in oscillating flame dynamics.

\section{Collective Behavior in Computing Networks: Emergence in Artificial Complex Systems}

Collective behaviors are found not only in fireflies, birds, or ants but also in human-built systems of sufficient interconnections and complexities, such as distributed computing environments, the internet, and even during the software evolution. Just as with natural systems, it becomes impossible to predict the system's evolution in complex digital systems on account of irreducibly large computation requirements. Thus, it becomes necessary rather than optional to talk of the collective behavior in its own right. On the other hand, discarding the details of constituent behavior make the system and complexity manageable. Furthermore, this approach also provides a good description of the system where different scenarios can be modeled and tested. For example, a new computing paradigm, called Emergent Computing, is under exploration where a high-level functionality at the global level arises in a non-linear manner from low-level interactions of system building blocks.

Similarly, in the context of complex algorithms, over the years, we have witnessed many examples of emergent behavior in cellular automata, artificial neural networks, and fuzzy motion controllers. For example, one of the most popular machine learning algorithms is Deep Leaning, which is based on artificial neural networks. While we can comprehend the logic of creating an artificial neuron or a small network of neurons, their behavior en-mass exhibits an additional level of complexity that is inexplicable, unpredictable, and usually not intended or programmed by the developer. This, too, exemplifies an emergent behavior of the system as a whole. At some level, this can be considered to be akin to the rise of human intelligence from the complex set of billions of neuronal connections.

\section{Emergence in Economic systems}

The grand macroeconomic landscape in general and the financial markets, in particular, are an example of a largely uncontrolled (although regulated) system where thousands of `free' individuals make decisions driven solely by their self-interests. The questions that occupy the minds of researchers in these settings are : how parts of a system give rise to collective behavior and how this system as a whole interacts with its environment. There is a large body of work that over the years has catered to these requirements: From the Invisible Hand theory of Adam Smith to treating financial markets as `Complex Systems,' the objective is to make predictions on the collective behavior, which in general is very different from a simple scaling up of the behavior of the individual agents~\cite{chen2002emergent}. In general, financial markets have all the traits to be considered as a self-organizing network of agents.

For example, many researchers have contemplated and proposed models on the emergent properties in wealth distributions. The simplest of these models is a model which thinks of economic agents like molecules in a gas that can exchange energy. A Macroeconomic behavior, then,  results from a large number of interacting agents, most of whom have only partial information about the state of the economy at that point in time~\cite{dosi2019more}. The complex interactions between the agents allow for the economy to evolve. What emerges in the aggregate may have little to do with what happens at the individual level. We can think of economic agents as sharing some common features: their interactions are spread out, the agents are different, and the agents can learn from their own experience as well as from the experiences of other agents they encounter. To deal with uncertainty, economic agents try to make sense of problems by making guesses, using past knowledge and experience. Because of this, agents continually update their internal decision-making model, adapting and replacing their strategies based on their experience. Just allowing them to do so ensures that the overall distribution of wealth in the economy follows an exponential distribution.

\section{Emergence in Physics}

Historically, reductionism has always been a dominant theme in physics, notions of which always get reflected in the assumption that reconciliation should exist between the phenomenological theories such as classical thermodynamics and hydrodynamics with more elemental theories such as statistical mechanics. Here, one is tempted to assume that a detailed understanding of the governing principles at the lower stratum should culminate in a complete understanding of the behaviors at higher strata.  However, emergence provides an alternative viewpoint and engenders a possibility where attempts of such reconciliation may not always be prolific. In the context of the mentioned example, thermodynamics can be considered an emergent reality manifesting itself in statistical systems. 

Additionally, in the modern physics literature, the focus on emergent phenomena also comes from the point of view of establishing tractable mechanisms and explanations for the macroscopic world of our experience without being consumed in overpowering details of the micro-constituents. This is extremely important for the rapid progress in applied sciences. In fact a closer look reveals that an entire domain, called Condensed Matter Physics, focuses on collective behavior and quantum many-body problems, hence, invariably is in search of emergent behavior. 
When we start our journey into condensed matter systems, we may naively conclude that since quantum mechanics provides a detailed and self-sufficient description of atoms, interaction of collection of atoms in terms of well-understood forces (such as Coulomb interaction among charged particles) should only be mechanistic exercise with no new intellectual challenges;  however, time and again this idea has been infringed upon. Today, the condensed matter systems (which are inherently large) are considered one of the most fertile grounds for witnessing instances where the emergent whole brings up unexpected macroscopic properties. The uncanny nature of collective electron behavior has brought about intense research interest in such systems and led to some of the most exotic discoveries such as the emergence of new kind of phases of matter (called topological matter, with aspects like fractional quantum Hall states), the emergence of Van der Waals forces by coarse-graining the interactions of neighboring molecules, observations of high-temperature superconductivity and a generalized understanding of phase transitions and critical behaviors. In the same spirit, the focus of the following discussion has also been kept on some of the examples of emergent phenomena in condensed matter systems.

\subsection {Phase Transitions as Emergent Phenomena}

Over the late twentieth century, it had been endured that the emergent phenomena in condensed matter systems invariably lead to new phases of matter and hence, serves as a cradle for neoteric physics. While being made up of identical microscopic constituents, these phases exhibit stark differences in their macroscopic properties. This necessitates the need to develop an apparatus that can explain how the emergent whole makes a transition from one macroscopic state to another, making the theory of phase transitions an elemental theme in investigations of condensed matter systems. Indeed, establishing the existence of phase transitions and fathoming their characteristics forms an indispensable subject matter in statistical and condensed matter physics. 

The phase of a system (state of matter in which its physical properties are uniform on a macroscopic length scale) is defined in terms of a thermodynamic function, typically Helmholtz or Gibbs free energy or partition function. The parametric dependence of the system's free energy on macroscopic system parameters leads to the possibility of change in the phase of the system, which is generally represented on a phase diagram (a parametric space where each point represents a phase of the system).  A representative example of a phase diagram is presented in Fig. 5, which exhibits specific features like phase boundaries, a triple point, and a critical point. Within a particular phase of matter, the partition function remains analytic; however, it experiences some form of the branch cut along the phase boundaries. Consequently, movement across the phase boundaries forces the system to experience non-analyticities, leading to phase transitions, which can be driven by: temperature, pressure, electric or magnetic fields. Conversely, we can also say that phase transitions entail an abrupt change in the system response functions; hence they correspond to the singularities of thermodynamic function~\cite{goldenfeld2018lectures,domb2000phase}.

From the perspective of physical origins, the phase transitions can be categorized as: (i) Classical or Thermal Phase transitions; (ii) Quantum Phase transitions. While classical phase transitions are driven by thermal fluctuations, quantum phase transitions at absolute zero temperature are initiated by variation in some non-thermal control parameters (such as magnetic field). A more meticulous classification method is provided in terms of the degree of singularity in the physical quantities. Thus, the discontinuous or first-order transitions have the first-order derivative of free energy itself to be discontinuous. This leads to a discontinuity in thermodynamic quantities such as enthalpy,  specific heat, entropy, and internal energy; an example being the melting of a three-dimensional solid and subcritical gas to liquid condensation. In contrast, quantities behave smoothly at second-order (or continuous) phase transition; in other words, only the second or higher-order derivatives of free energy are discontinuous. This prohibits discontinuity in thermodynamic quantities, and usually, their first-order derivatives are discontinuous. Examples of a second-order transition are the  liquid-gas transition around critical and supercritical points and superfluid transitions.

Depending on the region of operation, the same material can exhibit both a first-order or a second-order transition; for example, paramagnetic–ferromagnetic transition in magnetic materials around the Curie temperature ($T_C$) is second-order transition while for $T<T_C$ the magnetization in the same material exhibits a discontinuous transition when the external magnetic field is scanned from positive to negative values.

\begin{figure}
\begin{center}
  \includegraphics[width=0.7\textwidth]{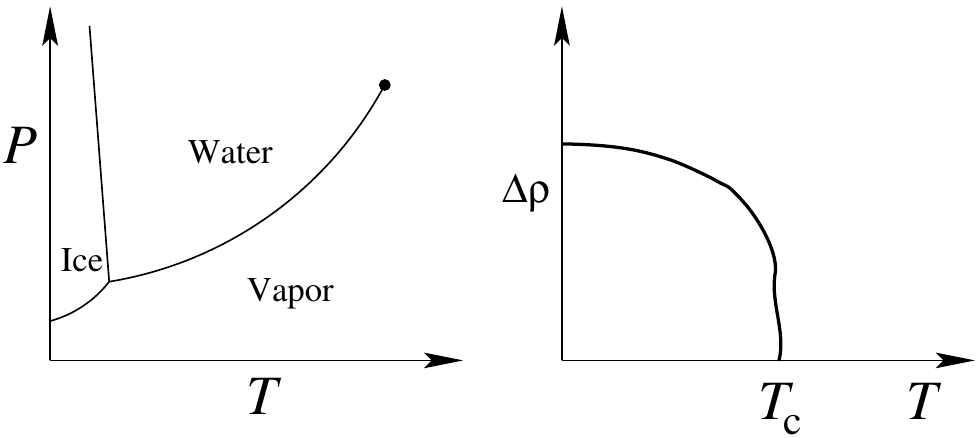}
  \caption{Phases of water (left), and the variation of order parameter near the critical point (right). The region $T < T_c$ corresponds to the symmetry broken phase where the order parameter $\Delta \rho \neq 0$ (here $\Delta \rho$ is nothing but the change in density away from the critical point).}
\end{center}
\end{figure}

\subsubsection{Discontinuous or First-Order Phase Transition}

Discontinuous phase transitions mark an abrupt change in thermodynamic properties with smooth change in system parameters (like temperature and pressure). Theoretically, they are described as the emergence of non-analyticities (singularities) in the thermodynamic function, with the phase transition point being marked in the parametric space where the thermodynamic potential becomes singular, exhibiting a jump or divergence. We will discuss two specific examples of such transitions: The first one being the gas to the crystalline phase transition. When a disordered gas makes a transition to the crystalline solid phase, a long-range order or the periodic arrangement of molecules gets developed in the system~\cite{sun2017crystallization}. It can be parametrized in terms of mass density $\rho(\boldsymbol{r})$ (which can also be considered to be an order parameter for the system)

\begin{equation}
    \rho(\boldsymbol{r})=\rho_{mean} + \sum_{\boldsymbol{k}\epsilon{\boldsymbol{G}}} [\rho(\boldsymbol{k}){e^{-i\boldsymbol{k}.\boldsymbol{r}}}+ C.C.]
\end{equation}

where the crystal structure is characterized by $\boldsymbol{G}$ (the set of reciprocal lattice vectors). When a temperature-driven phase transition is brought about by cooling the gas to the ordered crystalline phase, the translation invariance gets spontaneously broken, which gets reflected in the appearance of Fourier Modes $\rho(\boldsymbol{G})$. Experimentally it leads to the appearance of Bragg peaks in scattering. Usually, the emergence of these Fourier modes is discontinuous, characterizing a first-order transition. The second example we discuss corresponds to Liquid Crystals. The isotropic phase to nematic phase transition in liquid crystal is also a discontinuous or first-order transition. At high temperature, the rod-like molecules in liquid crystal solution are isotropically distributed; however, when cooled down below a particular phase transition temperature, an orientational order sets in where a combination of enthalpic and extended volume effects stimulate a spontaneous alignment (of rods) along a particular direction, called as the director or optic axis of liquid crystal~\cite{chester2013phase}. In quantitative terms, an orientational order parameter (based on anisotropy of distribution of rods) is defined in terms of the average of the second Legendre polynomial:

\begin{equation}
    S= \langle P_2(cos\theta) \rangle = \frac{1}{2}\langle {3~cos^2\theta-1} \rangle
\end{equation}

where $\theta$ is the angle between the director and the local orientation of the rod and the average $\langle . \rangle$ is performed over the equilibrium distribution of all the rods. At the transition temperature, the scalar order parameter $S$ exhibits a discontinuity. Similarly, the subcritical liquid to gas transitions also exhibits first order characteristics.

\subsubsection{Anomalous behavior: Critical Phenomena and Continuous Phase Transitions (Higher Order Phase Transitions)}

Continuous phase transitions manifest in the vicinity of the critical point of the system (when phase boundaries get obliviated) and are characterized by indistinguishability between the two phases~\cite{uzunov1993introduction}. The pedagogical example of a continuous phase transition is the supercritical liquid to gas transition (which is a second-order phase transition). Another representative example is the ferromagnetic to paramagnetic transition in magnetic materials (in the absence of external field) around the Curie temperature (or critical point). In these scenarios, the magnetization in the material constitutes the order parameter, which is defined as

\begin{equation}
    \boldsymbol{M}= \frac{1}{N} \sum_{i} \langle \boldsymbol{S}_i \rangle
\end{equation}

where $\boldsymbol{S}_i$ represents the spin corresponding to the individual spin element, and $N$ represents the density of these elements. The magnetization $\boldsymbol{M}$ grows smoothly from zero in case of $T>T_C$ (paramagnetic phase) to $T<T_C$ (ferromagnetic phase). Thus, a spontaneous magnetization gets developed in the material (which is a smooth function) while undergoing the phase transition from a disordered (high temperature) phase to an ordered (low-temperature phase). This is an example of second-order phase transition.

\subsubsection{Spontaneous Symmetry Breaking in Phase Transitions}

Symmetry breaking refers to the idea that a condensed phase possesses lesser symmetries than its uncondensed peer. Spontaneous symmetry breaking leads to the manifestation of phases of reduced symmetry while trespassing from disordered phase~\cite{sinha2016explicit} (the so developed ordered phase possesses lower symmetry than the system itself). Alternatively, spontaneous symmetry breaking is said to have taken place whenever the Hamiltonian of the system possesses a symmetry that is not manifested by the ground state (the state of minimum energy) of the system, which is an exotic example of emergent behavior. The phenomenon is associated with the phase transitions happening between the two realizations of different symmetry properties and can be thought of as a remnant occurrence (arising from an underlying symmetry), which, although is not manifested in the macroscopic phase properties, but gets reinstated upon increasing the energy of the system. 

As implied before, more often than not, phase transitions entail the development of some order in the thermodynamic system. This arbitrary inclination towards order causes some or the other symmetry of the system to become veiled as the transition progresses. A quantitative description of this phenomenon is provided by the construct of an order parameter (a quantitative measure of the degree of asymmetry in the broken-symmetry phase), which becomes larger and larger as we pierce deeper into the ordered phase. In the case of thermally driven phase transitions, the high-temperature (disordered) phase exhibits higher symmetry than the low-temperature (ordered) phase. The earlier mentioned ferromagnetic to paramagnetic transition can also be explained from a symmetry point of view : Considering only the spin-spin interaction, the system Hamiltonian of a ferromagnet can be represented as
\begin{equation}
    \boldsymbol{H}= \sum_{ij} J_{ij} \boldsymbol{S}_i\boldsymbol{S}_j
\end{equation}

This Hamiltonian is invariant under rotation operation. When $T>T_C$, the ground state also possesses this symmetry, i.e., there is no preferred alignment in the system (called the paramagnetic phase). However, if the system is cooled to temperatures such that $T<T_C$, then the ground state loses the rotational symmetry (spin rotation symmetry is broken or becomes hidden), and a preferred direction of alignment is randomly chosen by the system (called the ferromagnetic phase). 

It is important to note that while spontaneous symmetry breaking always engenders a phase transition, the reverse need not be true. Many instances have been observed where no (apparent) symmetries of the system are being broken during the phase transition event (such transitions can only be first-order).

\subsubsection{An Exactly Solvable Thermodynamic System: Ising Model}

Ising model is an analytically solvable microscopic mathematical model which can be employed to study how the interaction between the elements can bring about emergent phenomena such as phase transitions of varying origins and characteristics~\cite{huang2009introduction}. It provides a first-hand experience of Universal behavior (which is a feature of Critical phenomena) where a common theory is capable of explaining the phase transitions occurring in liquids, gases, magnets, or even in superconductors. In the perspective of the model to be representing magnets, it comprises of discrete variables which represent atomic spins (or Ising spin $\boldsymbol{S}_i$), present in one of two possible states (+1 or -1). The Hamiltonian for spin-$\frac{1}{2}$ Ising model is defined as:

\begin{equation}
    \boldsymbol{H}= -\sum_{\langle ij \rangle} J_{ij} \boldsymbol{S}_i\boldsymbol{S}_j-\sum_{i}  \boldsymbol{h_i}\boldsymbol{S}_i
\end{equation}

\begin{figure}
\begin{center}
  \includegraphics[width=0.33\textwidth]{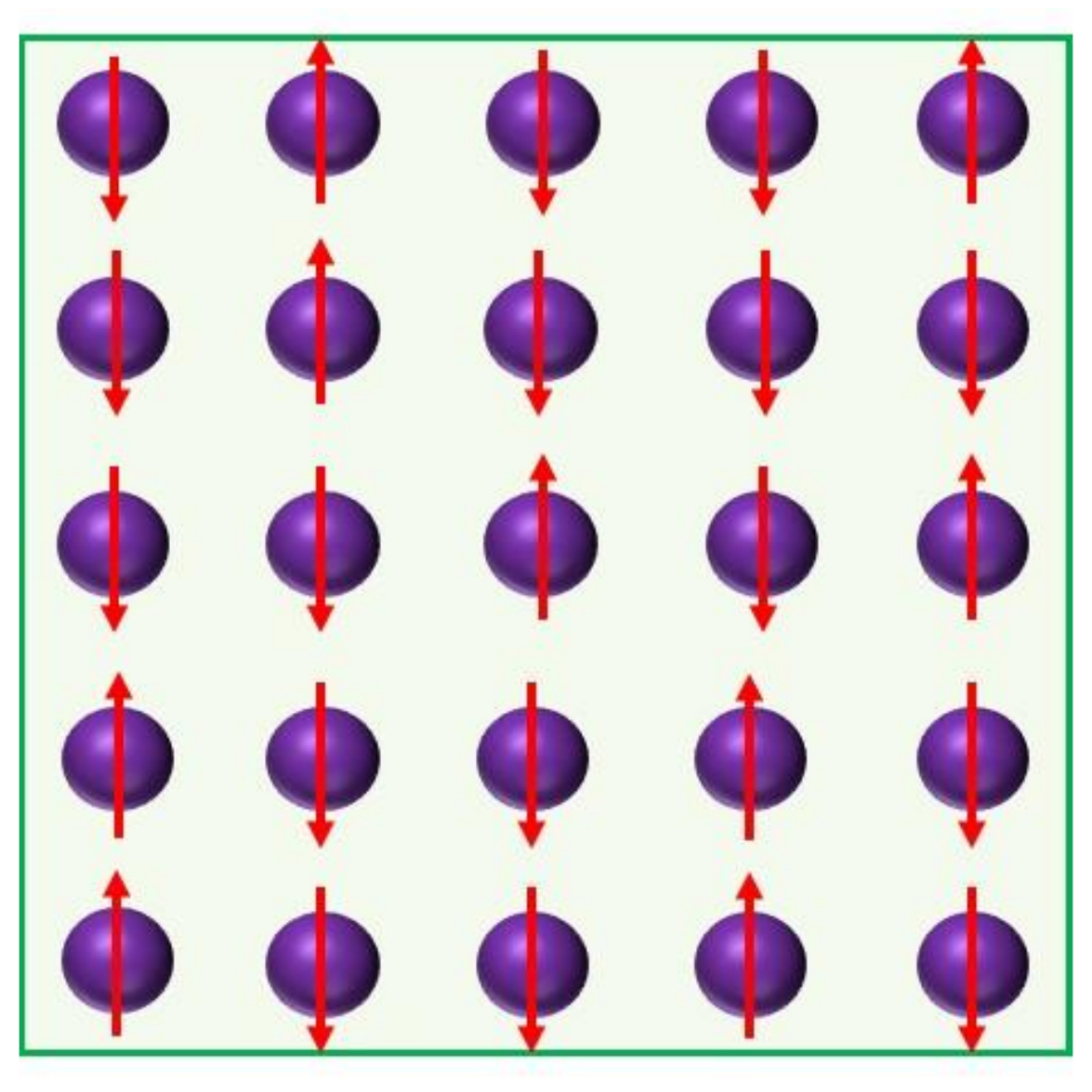}
  \caption{Architecture for the two-dimensional square lattice Ising model.}
\end{center}
\end{figure}
 here $\boldsymbol{h_i}$ is the local external magnetic field (expressed in energy units). The Hamiltonian comprises of two terms, the first being the interaction term between neighboring spins; it is noteworthy that only nearest-neighbor interactions are considered. $J_{ij}$ represents the coupling constant or exchange interaction, the magnitude of which represents the interaction strength, while its sign represents the tendency of spins to align or anti-align. In case of $J_{ij}>0$, the interaction is ferromagnetic (aligned) while for , the interaction is $J_{ij}<0$ anti-ferromagnetic (anti-aligned). The second term represents the spin interaction with the external magnetic field. The dynamics of such Ising model is governed by the fact that although the system wants to stay in the lowest energy state but the external perturbations disturb the system and engender the possibility of different structural phases. Fig. 6 shows an example where Ising spins are arranged in a two-dimensional square lattice. 

Indeed, this two-dimensional Ising model is the simplest statistical model to manifest the phase transitions(the one-dimensional Ising model is devoid of phase transitions). Although it only provides a simplified description of the macroscopic reality, it is capable of exhibiting phase transitions as well as critical phenomena, which makes it a very useful theoretical tool for gaining crucial insights. For example, the Ising model can predict the ferromagnetic to paramagnetic phase transition in terms of $Z_2$ symmetry breaking.

\subsubsection{Landau's Generalized Theory of Second Order Phase Transitions}

The Russian physicist and Nobel laureate Lev Landau's work in the 1940s established the foundations for how we think about phase transitions today. Landau's theory of phase transition has been one of the cornerstones of twentieth-century physics and proved to be particularly instrumental in unearthing various effects related to magnetism and superconductivity~\cite{landau2013course}. Specifically, the theory was introduced to universally understand the continuous (or second-order) phase transitions.  The characteristic feature of Landau's theory is that it is a phenomenological theory. Consequently, unlike a microscopic theory, it does not concern itself with atomic-level details to get to the macroscopic governing laws. Instead, it has been argued that the essential physics of phase transitions can still be captured by working at a level of obscurity. In this spirit, an emergentist viewpoint on phase transition also gets vindicated. The central idea of Landau's theory is that of an order parameter; in fact, it can be treated as an effective theory of the order parameter. The procedure lies in approximating the free energy of the system such that the non-analyticities corresponding to the phase transition can be incorporated. For this, first of all, an order parameter ($\psi$) is appropriately defined for the process. Then based on symmetry considerations alone, the free energy is defined in terms of a Landau Functional ($\Delta{F(T,\psi)}$), which is a function of $\psi$ and expanded as a Taylor series in terms of $\psi$. 

\begin{equation}
    F= F_0(T)+ \Delta{F(T, \psi)}
\end{equation}

\begin{equation}
   \Delta{F(\psi)}= a+ b\psi^2+c\psi^4
\end{equation}

Thereafter a minimization of Landau Functional is performed over $\psi$ as a function of temperature. At this level, the system is described to have a mean state; thus, Landau's theory is a mean-field formulation.  It turns out that the primary consideration in Landau's theory are the symmetries possessed by the system, all of which are reflected in the Landau Functional as well. Thus, the theory applies only to phase transitions which also induce a symmetry change. Landau's framework is also known to provide a glimpse of the notion of Universality in physics as two systems exhibiting the same symmetries invariably display similar characteristics in Landau's theory (even though these systems may belong to two entirely disjoint domains and hence, possess disjoint macroscopic properties).
\\

Having pondered upon the phase transition phenomena in quite some detail, we move on to discuss some of the other instances of emergent behavior in condensed matter systems.

\subsection{Geometric Phases and Emergent Fields}

 In quantum many-body systems, many a time, the eigenfunctions of Hamiltonian are classified into disjoint subgroups (like spin polarization-based classification in Mott insulators); electron wavefunctions in these subgroups span only a subspace of Hilbert space which gives rise to a nontrivial geometrical structure. 
In his seminal work of 1984, Berry showed the intriguing consequences that such settings could carve onto the evolution of wavefunction in these subspaces. Specifically, when a system adiabatically evolves around a closed path in this parametric subspace, geometric phases arise~\cite{berry1984quantal}. The associated Berry Connection gives rise to an effective gauge field for the quantum system, called the Emergent Electromagnetic Field. These emergent electromagnetic fields are presenting fledgling grounds for observing distinctive phenomena like non-Abelian gauge fields in condensed matter systems.

Consider a quantum  mechanical system with non-degenerate states and which evolves adiabatically in a parameter space $\boldsymbol{X=X}(t)$. In such a scenario, the solution of the time-dependent Schrodinger equation

\begin{equation}
    i\hbar \frac{\partial \psi}{\partial t}= \boldsymbol{H(X(t))} \psi(t)
\end{equation}
is given by
\begin{equation}
    \ket{\psi(t)}=\sum_j a_j(t_0) e^{i\phi_j(L)} e^{-\frac{i}{\hbar} \int_{0}^{t} \,dt' \epsilon_j(\boldsymbol{X(t')})} \ket{\psi_j(t)}
\end{equation}
here $\ket{\psi_j(t)}$ and $\epsilon_j(\boldsymbol{X}(t'))$ are the time evolved eigenstates and the eigenvalues respectively of the Hamiltonian $\boldsymbol{H(X(t))}$ and $\phi_j(L)$ denotes the Berry phase. The initial state of the system is defined from eqigenvalue equation as $\ket{\psi(t_0)}=\sum_j a_j(t_0) \ket{\psi_j(t_0)}$. The expression of the Berry phase is given in terms of the Berry Connection as:

\begin{equation}
     \phi_j(L) =\oint_L \boldsymbol{A_j(X)}. \,d\boldsymbol{X}
\end{equation}
and the Berry Connection is defined as

\begin{equation}
    \boldsymbol{A}_{j}(\boldsymbol{X}) =i\expval{\nabla_{X}}{\psi_j}
\end{equation}

\begin{figure}
\begin{center}
  \includegraphics[width=0.37\textwidth, angle=00]{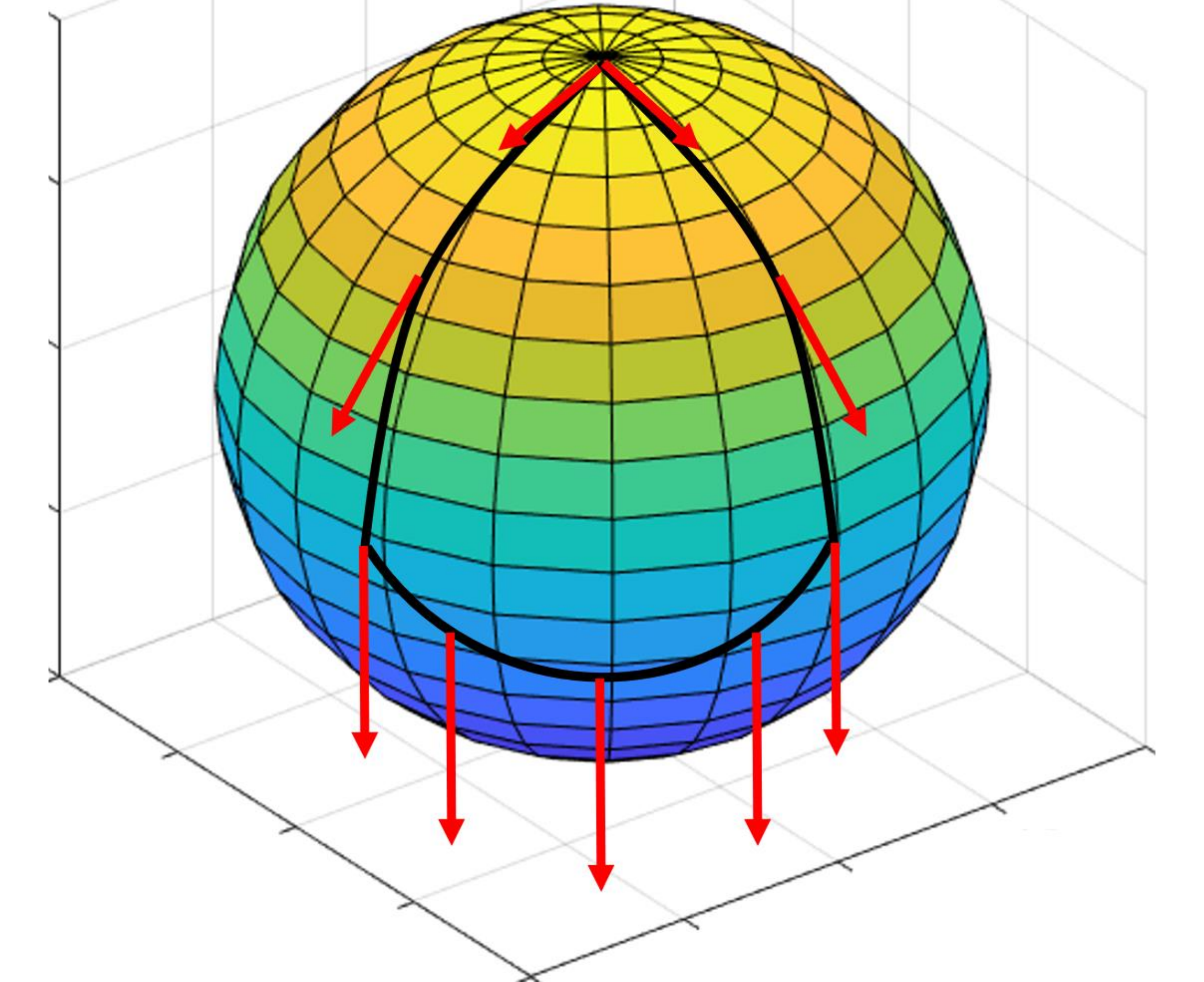}
  
  \caption{Parallel transport of a vector along a closed contour leads to gauge independent Geometric phase.}
\end{center}
\end{figure}

Berry connection is a gauge-dependent vector-valued object, which acts as a vector potential in $\boldsymbol{X}$ space. Therefore, we can define emergent electromagnetic fields using this object. 

The experimental manifestations of these emergent fields can be observed in the settings of Anomalous Hall Effect~\cite{nagaosa2010anomalous}. In these scenarios, the emergent magnetic field gives rise to a transverse acceleration (with respect to the electric force or current). This velocity is called Anomalous velocity, and it engenders an Anomalous Hall Effect without the requirement of Lorentz force.

\subsection{Macroscopic Quantum Effects}

Quantum phenomena are said to be macroscopic when the associated quantum states are occupied by a large number of particles. This does not look very familiar as we have this notion ingrained in us that the quantum effects are observable only at atomic scales; however, in many instances, these effects do manifests themselves in macroscopic settings as well. Particularly, in the case of condensed matter systems and other quantum many-body systems, many novel phases of matter have been discovered, whose characteristics are quite surprising and can only be explained in terms of effects that are unique to quantum mechanics. The examples include but are not limited to  superconductivity, quantum Hall effect, Bose-Einstein Condensation, and superfluidity~\cite{kruchinin2021modern,thouless1982quantized,bogoliubov1970lectures,feynman1957superfluidity,tilley1990superfluidity}. Here, we will describe two of the fundamentally related phenomena, superconductivity, and superfluidity. A combined definition of these can be given as: the flow of matter or electric charges without friction. These phases and their associated theories provide a unique exhibition of amalgamation of fundamental quantum mechanical principles and macroscopic descriptions of bulk material (in terms of emergent behavior).

\subsubsection{Superconductivity}
The phenomena of superconductivity refer to a state when the electrical resistance of a material vanishes and a corresponding expulsion of magnetic field lines (Meissner effect) takes place~\cite{kruchinin2021modern}. Unlike conventional metallic conductors whose resistance keeps on decreasing with the reduction in temperature (but remains finite even at near absolute zero temperatures), the superconductors exhibit a critical temperature below which the resistivity abruptly drops to zero. The onset of superconductivity is accompanied by a phase transition. Some of the prominent superconductors are aluminum, niobium, magnesium diboride, and yttrium barium copper oxide. Based on their response towards the applied magnetic field, the superconductors are classified as Type I or Type II (high temperature) superconductors. 

Superconductivity is a very pertinent example of emergent phenomena as most of the explanations of superconductivity subscribe to some novel collective behavior of electrons such as the formation of Cooper pairs~\cite{cooper1956bound} and their condensation or strong correlation of electrons in certain materials. Thus, the cooperative behavior of electrons, when understood in light of quantum effects, explains this novel behavior as the realization of a macroscopic condensate.

\setlength{\fboxsep}{10pt}
\begin{center}
\fbox{
  \parbox{0.8\linewidth}{
{\bf Bose-Einstein Condensate (BEC) :} It is a quantum-many-body phenomenon that engenders an exotic state of matter. When a dilute gas is cooled to around a near absolute zero temperature, a large fraction of atoms of the gas jump to the ground state and collapses into a single wave function, and form a monolithic object. This is a representative example of macroscopic quantum mechanical phenomena, where the wavefunction interference becomes apparent at macroscopic scales. It turns out that the formation of BEC is also a phase transition event that occurs below a critical temperature. To estimate this temperature, let us assume that there are $n$ atoms in a unit volume. The separation between atoms $l \sim 1/n^{1/3}$. If this separation is less than the de Broglie wavelength $\lambda_{dB} = h/p$, where $h = 6.62 \times 10^{-34} Js$ is the Planck's constant and $p$ is the momentum, the wave functions of the atoms overlap with each other. The kinetic energy $p^2/2m$ of atoms at temperature $T$ is $k_B T$. So that $T = h^2/2mk_B\lambda^2_{dB} \sim (h^2/2mk_B) n^{2/3}$. At this temperature, it starts forming a Bose-Einstein condensate.

  }
}
\end{center}

The first successful description of superconductivity was the Bardeen–Cooper–Schrieffer theory (BCS theory)~\cite{bardeen1957microscopic}, which explains the properties of Type I superconductors. According to this framework, superconductivity is a macroscopic effect that can be attributed to the condensation of Cooper pairs (electron-phonon interaction). Although electrons are fermions, the formation of Cooper pairs accredits them with some bosonic properties. This engenders the possibility of Bose-Einstein condensation~\cite{bogoliubov1970lectures} at very low temperatures, and hence the superconducting state is also termed as a single-valued quantum state. Although BCS theory provides the foundations to understand superconductivity and has been successful with Type I superconductors, not all experimental observations can be explained by it. Particularly, in the last few decades, many instances of unconventional superconducting behavior have been noticed where some non-phononic mechanism of superconductivity is at work. It all started in 1986 with the discovery of high-temperature superconductivity in La$_{2-x}$Fr$_x$Cu$_2$O$_4$ ~\cite{bednorz1986possible}. These materials are insulators with a very large repulsion between electrons that prevents them from double occupancy. In these materials, electrons are strongly correlated, which also can lead to superconducting behavior, termed as electron correlation mechanism of high-temperature superconductivity (enunciated by Anderson and coworkers~\cite{anderson2012hightc,anderson2004physics}).

Another interesting property associated with superconductors is that of magnetic flux quantization. It has been observed that around a loop or hole inside a bulk superconductor, the flux traversing through the area is quantized. The unit of quantization of magnetic flux is termed as superconducting magnetic flux quantum ($\Phi_0$), which is itself a physical constant. Its value is independent of the geometry of the loop and also independent of the material as long as it is in the superconducting state. The inverse of flux quantum is called Josephson constant ($K_J=\frac{1}{\Phi_0}$).

\subsubsection{Superfluidity}

In simplest terms, we can define superfluidity as a kind of superconducting state of neutral particles. In more formal language, it is defined as a state of a fluid with zero viscosity so that it can flow without loss of kinetic energy~\cite{feynman1957superfluidity,tilley1990superfluidity}. The primary requirement of observing superfluidity is that we must have access to a material that does not freeze even at near absolute zero temperatures, a relatively uncommon characteristic but found in Helium. Helium is a weakly interacting gas and is too difficult to liquify. Heike Kamerlingh Onnes in 1908 made a major breakthrough when he could chill helium gas to 4.2 degrees above absolute zero to form a liquid at 1 atm pressure (it immediately lead to the discovery of superconductivity in mercury).  However, in 1938 it was discovered that strange properties emerge in liquid helium (helium-4) when it is further cooled to 2.17 K; it suddenly starts exhibiting a superfluid behavior. The superfluid helium phase is called the He-II while the ordinary liquid phase is called He-I, the transition between them is called the $\lambda$-transition (the specific heat at the $\lambda$ point diverges). The thermal conductivity of superfluid He-II increases with decreasing temperature. 
\begin{figure}
\begin{center}
\includegraphics[width=0.5\textwidth]{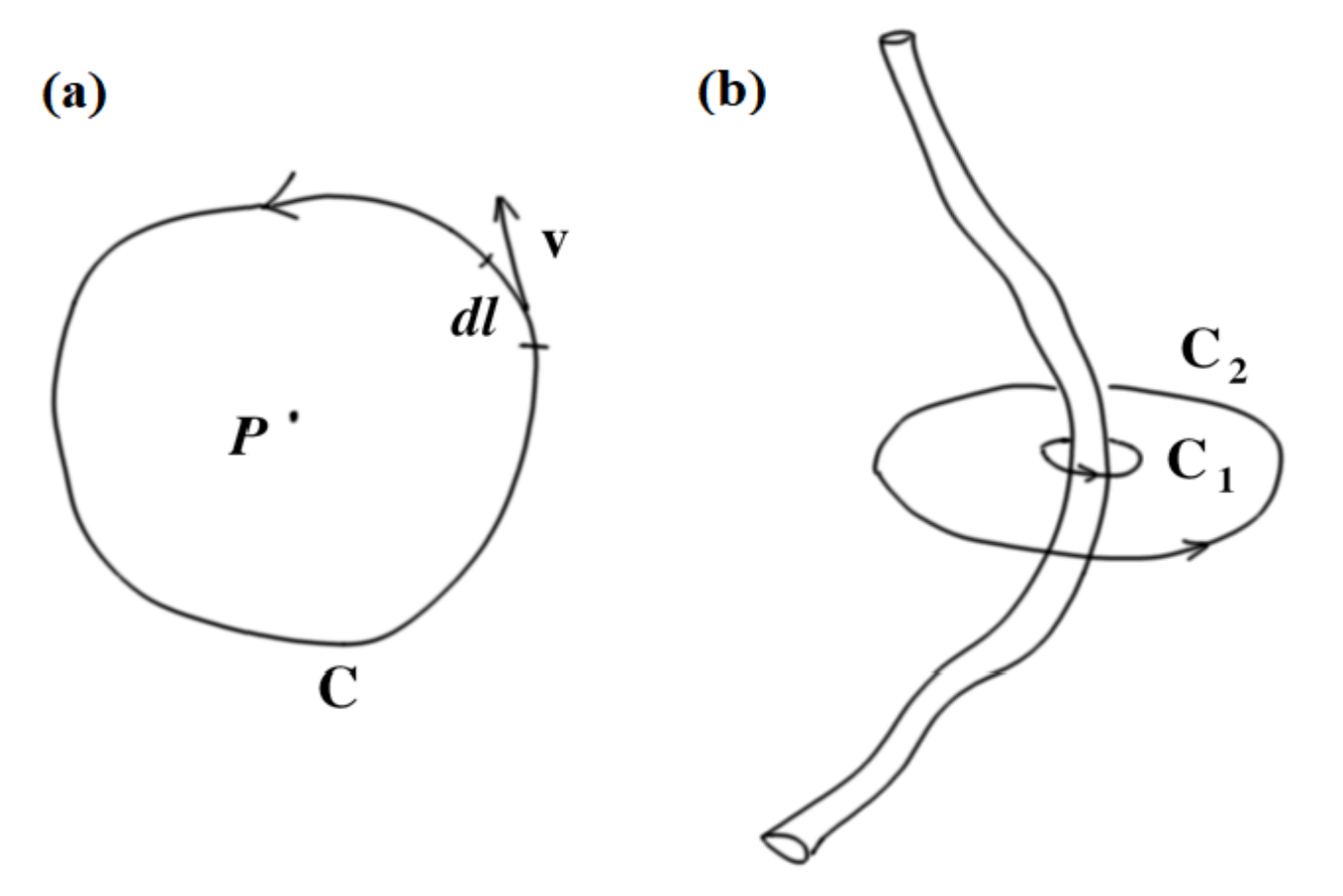}
\caption{The quantum vortex and circulation around a vortex tube.}
\label{vortex}
\end{center}
\end{figure}
The theoretical explanations of the phenomena were provided in terms of the formation of a monolithic BEC. Atoms of helium-4 behave as bosonic particles with integer spin values; hence at very low temperatures they can make a transition to the BEC state. Fritz London used the theory of Bose-Einstein condensate for ideal gas and estimated  that at the density of He-II, the transition temperature would be around $3 K$. Landau later proposed a two-fluid theory in which the He-II phase has a normal component and a superfluid component. But the story about the superfluidity in helium does not end here; after many years quite unexpectedly, the superfluid properties were also found in helium-3. Helium-3, the lighter of isotopes of helium, liquifies at 3.2 K. Furthermore, it does not show superfluid behavior at these temperatures. The reason being that helium-3 atoms behave as fermions and follow a different statistics. However, fascinating things start happening when helium-3 is cooled to extremely low temperatures of 2.491 mK when it starts behaving as a superfluid. At this temperature, helium-3 atoms form pairs and behave as bosons; this mechanism is very similar to the electron pairing mechanism already explained in the context of superconductivity. 
\setlength{\fboxsep}{10pt}
\begin{center}
\fbox{
  \parbox{0.8\linewidth}{
    {\bf The Quantum Vortex:} Vorticity is a measure of rotation in a fluid. A flow pattern is called irrotaional
    at a point $P$ if the circulation $\Gamma = \oint_C \bm{v} \cdot d\bm{l}$ vanishes, where the closed contour $C$ encloses the
    point $P$. Vortices also form in superfluids with an extra feature that they are quantized,
    \begin{align}
      \oint_C \bm{v} \cdot d \bm{l} = \frac{h}{m}n,
    \end{align}
    where $n$ can take integer values. This simply means that vorticity is a conserved quantity. This is demanded by
    the single valuedness of the wave function $\psi(\bm{r})$. Consider for instance the closed contour
    $C$ in \fig{vortex}(a). We can parametrize  $d\bm{l} = d\bm{l}(\theta)$ as a function of angle $\theta$
    such that the contour integral corresponds to integrating over $0$ to $2\pi$. The velocity can be written as
    $\bm{v} \sim \psi^* \bm{p} \psi/m$. But since $\psi$ needs to match at $\theta=0$ and $2 \pi$ the contour integral
    should give  an integral multiple of $2 \pi*\hbar$, the quantum of angular momentum.
  }
}
\end{center}

Among many strange properties of superfluids, the one which gets enormous attention is the formation of quantum vortices in superfluids. Indeed it has been observed that the vortices in superfluids are quantized. Such behavior can only be explained by quantum mechanics, where the whole system is described by a macroscopic wave function, and it is fundamentally similar to the flux quantization observed in superconductors. Consequently, these aspects are studied together as the formation of a quantum vortex, representing a quantized flux circulation of a physical quantity.    

\setlength{\fboxsep}{10pt}
\begin{center}
\fbox{
  \parbox{0.8\linewidth}{
    {\bf Flux Quantization in Superfluids and Superconductors}: To demonstrate the quantization of flux, let us define the wave function to be 
    $\psi \simeq \sqrt{n} \exp(\mathrm{i}\theta)$, where $n$ is the number density of Cooper pairs, and $\theta$ be a global phase (which persists because of the rigidity acquired by the wavefunction in the superconducting state). Now we consider a superconducting ring or torus. The charges in this ring are rotating around the infinite vortex, and the resulting current density is given by
    $\bm{j} = q(\psi^*(\bm{p} - q \bm{A})\psi - \psi(\bm{p} - q \bm{A})\psi^*)$  
    or $\bm{j} = (q n \hbar/m) \nabla \theta - (q^2n/m) \bm{A}$. Taking contour integral (around some arbitrary path $C$)
    on both sides we have $\oint \nabla \theta \cdot d\bm{l} = 2 \pi s$, where $s$ is an integer that represents the difference in the phases
    of the wave function and $\oint \bm{A} \cdot d\bm{l} = \Phi$ is the magnetic flux enclosed by the contour.
    We may therefore write
    \begin{align}
      \Phi + \frac{m}{qn} \oint_C \bm{j} \cdot d\bm{l} = \frac{2 \pi s \hbar}{q}. 
    \end{align}
    If the contour $C$ is close to the surface of the superconductor, currents will be non-zero, making the sum of flux enclosed
    and the circulation of current to be quantized. However, when the contour is deep inside the superconductor, the current density $\bm{j}$ vanishes as per the Meissner effect, and we have $\Phi = 2 \pi\hbar s/q$. Therefore, the magnetic flux through the superconducting ring turns out to be quantized with the flux quantum being $\Phi_0=2 \pi\hbar/q$. It was experimentally found that the charge $q=-2e$ which leads to a flux quantum of $\Phi_0 = h/2e \simeq 2.0678 \times 10^{-15} \text{tesla m}^2$.
    
  }
}
\end{center}

\subsection{Quasiparticle based Formalism as Emergent Phenomena}

Given the Hamiltonian of a quantum system, which incorporates electron-electron and electron-ion interactions, a variety of emergent phenomena can arise. At times, the sheer volume and complexity of these interactions make carrying out simulations to be an impractical task, let alone drawing rational conclusions out of it (this is particularly true for condensed matter systems where the number of particles is of the order of $10^{23}$ with strong interactions).  A very prolific and efficient description of such phenomena is provided by the concept of Emergent Quasiparticles~\cite{simula2019quantised}. Quasiparticles and collective excitations emerge from the complicated interactions occurring between a large number of fundamental particles, and they are considered to be weakly interacting. It was Landau who first proposed a Quasiparticle-based description to promulgate an efficient and working description of microscopically complex phenomena.
\begin{table}
  \begin{tabular}{|l|p{11.5cm}|}
    \hline
        {\bf Quasiparticle}  & {\bf Brief description} \\ \hline
        Magnetic Skyrmion & {Topologically protected spin textures manifesting themselves in magnetic systems. Specified in terms of topological charge or topological quantum number. Useful in spintronics.} \\ \hline
        Exciton  & Bound state of electron and hole. Carries no charge but may carry spin. \\ \hline
        Plasmon & Quantum of plasma oscillation or coherent excitation of plasma \\ \hline
        Phonon  & Quantum of structural vibrations or collective excitation of atomic vibrations. \\ \hline
        Magnon  & Quantized spin waves in crystals. Bosonic quasiparticles with spin-1. \\ \hline
        Polariton  & Bosonic particle resulting from strong coupling between EM waves and electric or magnetic dipole excitations. Manifestations of avoided crossing. Plasmon polaritons, exciton polaritons, phonon polaritons. \\ \hline
        Anyon  & Occurs in two dimension systems. Neither behaves as fermion nor as boson, obeys fractional statistics.  \\ \hline
        Polaron  & Bound state between electron and atoms in a dielectric material (electron-phonon coupling) or dressing of electrons by phonons \\ \hline
        Holon, Spinon and Orbiton  & Electrons in solids can be described as a bound state of these quasiparticles. Spinons sustain the electrons spin, holons carry the charge and orbiton sustain the orbital location. \\ \hline
        
        Relativistic Fermion:  & Fermionic quasiparticles in the relativistic limits. \\ 
        1.Dirac Fermion  & Possess a non-zero mass and charge. \\
        2. Weyl Fermion  & Massless but exhibit a charge. \\
        3. Majorana Fermion  & Have mass but no charge. These are half an electron and half a hole in the same place and time. Carry zero energy and zero charge. Emerge in superconductors or quantum spin liquids.\\\hline
        
        Roton  & Occurs in superfluids due to long-range dipole interaction or spin orbit coupling. \\ \hline
        Soliton & Solitary wave packet that appears due to nonlinearity and dispersion. It is a self-reinforcing wave packet that maintains its shape while propagation. \\ \hline
        
        Magnetic Monoploes  &  Emerge in condensed matter systems such as spin ice, carry effective magnetic charge.\\ \hline
       
        Composite Fermions  & Bound state constituting of electron and even no. of quantized vortices (magnetic flux quanta). These are anyons which can have fractional charge. Employed to explain fractional quantum Hall effect.  \\ \hline
        
  \end{tabular}
  \caption{List of quasiparticles.}
  \label{tab}
\end{table}

 In this perspective, collective behavior in large and complicated quantum systems leads to the genesis of quasiparticles with novel and often strange features. Quasiparticles are characterized by their own well-defined properties such as mass, charge, and spin. A list of quasiparticles along with some of their recognizing features is summarized in Table  (\ref{tab}).

\subsection{Quantum Materials and the Future Outlook of Emergent Phenomena}

A major triumph of twentieth-century physics was the recognition of emergent phenomena in many-body quantum systems and the development of a prolific and successful explanation of it in terms of phase transitions and symmetry breaking. Starting in the 1980s, however, two crucial developments took place in condensed matter physics, which made it very clear that the landscape for emergent properties is much bigger than we foresee. The first one was the concept of topological order~\cite{wang2017topological} (which is a torchbearer for understanding phenomena like fractional quantum Hall effect), and the other was the discovery of novel physics in systems with strongly correlated electrons~\cite{senechal2006theoretical}, due to strong Coulombic interactions (which led to high-temperature superconductivity). This promulgated the requirement of a broader description of emergent phenomena, and the term Quantum Materials was coined with a specific focus on quantum emergence. The term also reflects the fact that a broader class of such phenomena are potentially crucial for the development of quantum technologies, the examples include topological superconductivity, Mott transitions and Doped Mott insulators for high-temperature superconductivity, real space Berry phase effects in Skyrmions systems, and Topological Hall effect, emergent magnetic monopoles and topological quantum computing using Majorana fermions~\cite{frolov2020topological,phillips2020exact,fert2017magnetic,kurumaji2019skyrmion,lian2018topological}. From these considerations the search of emergent phenomena in quantum materials would invoke a convergence of principles from various fields of research such as Spintronics, Mottronics, Quantum topology, and Topological electronics.

\section{Emergence in Biological Systems}

The realm of biology is concerned with the ``living" organisms, particularly with their functioning and interaction dynamics. Although significant strides of progress have been made in the domain, one perennial question that remains unanswered is about the origins of life itself. How does a collection of otherwise inanimate molecules organize together to form a living entity that can shape its environment? The answer seems distant, but over the years, it has been realized that the origin of life can be studied in the purview of emergent phenomena, as, at each level of organization, a new and unprecedented level of functionality arises. Yes, perhaps the most striking and intriguing emergent phenomena of all is that of life!

Although a complete understanding of the emergence of life processes is not yet in sight, many significant proposals have been made which familiarize us with some or the other intermediary process. For example, one proposal assigns the origination of life to the ability of RNA molecules to catalyze chemical reactions and self-replicate. RNA enzymes (ribozymes) can catalyze (start or accelerate) chemical reactions that are critical for life~\cite{alberts2003molecular}. One of the most critical components of cells, the ribosome, is composed primarily of RNA. If the RNA world existed, it was probably followed by an age characterized by the evolution of ribonucleoproteins (RNP world), which in turn would have ushered in the era of DNA and longer proteins. DNA has better stability and durability than RNA; this may explain why it became the predominant storage molecule. Protein enzymes might have come to replace RNA-based ribozymes as biocatalysts because their greater abundance and diversity of monomers make them more versatile. 
Even if we set aside the question of providing evidence of such a hypothesis, our difficulties still do not get trimmed. The question of how inanimate, simple chemicals came to encapsulate themselves into a bag called a vesicle, thereby sequestering and protecting these chemicals from the outside world, is equally vital in fathoming the origins of life. In this context, it is generally agreed that two requirements are central. The first is that the constituents of life should be protected from their external environment; however, at the same time, an efficient passage for nutrients must be given. The second is that these materials must carry the requisite structure to allow their replication. Having said that, we still have no answer to any of the Whys and Hows here as well. All we can say is that the ability of cells to form a protective lipid coat around themselves was undoubtedly pivotal in the origin of life, and the mineral-rich, warm, and wet surroundings of under-sea vents might have provided optimal conditions for it.

Further, the adventure does not end here; from single-celled ancestors to the multi-cellular structures that make up complex life was yet another giant leap that brought about new forms of complexity. This brings us to an even greater mystery associated with the emergence of Intelligence in living entities. Once life forms reached the point of multicellularity, a pivotal moment in evolution emerges where different cells can perform different functions while working in harmony, bringing up the question of some form of `intelligence.'  We have seen similar behavior in the context of Swarm Intelligence; however, in the present context, the sheer complexity and evolutional hierarchy is overwhelming. Still, can we pin down some meaningful connections? The pursuit of answering questions of this sort has actually led to the rise of many new disciplines which bring in the methods and approaches from other domains to establish a deeper understanding of life processes.

\subsection{Emergence in Biochemistry: Autocatalytic Sets and Origins of Life}

Autocatalytic sets are self-sustaining chemical plants, consisting of groups of molecules, where the outcomes of one reaction form the feedstock for the other~\cite{hordijk2018autocatalytic,hordijk2010autocatalytic}. This constitutes a self-contained cycle of reactions that behaves like an entity with its own characteristics. Starting with many such autocatalytic sets and with enough complexity, functionally closed structures may arise at a higher level, giving rise to emergent behavior. This provides a mechanism for realizing novel properties at various levels of organizations. The mathematics of such interactions is independent of the chemistry employed, and in general, independent of the nature of the building blocks; such features are typical to an emergent phenomenon.

\subsection{A Biophysics perspective on Emergent Phenomena}

Biophysics is a rapidly evolving discipline that applies the principles of physics to investigate biological phenomena. More accurately, it attempts to explain biological functioning in terms of physical properties and the dynamics of molecules. 
In recent years one playground for such  that has been gaining traction among the scientists is that of ``Active Matter." Active matter possesses the ability to exhibit collective behavior and hence can form spontaneous patterns~\cite{sanchez2012spontaneous}. The behavior is very similar to the natural emergent phenomena such as flocking of birds and structure-forming cytoskeletons of cells. Scientists have even synthesized the active matter in the lab using biological building blocks like microtubules. With this ability to emulate the collective behavior in the lab, the prospects of a detailed and comprehensive study are now wide open.

\subsection{Quantum Biology: A Glimpse of Future Research}

Over the last few decades, another new interdisciplinary domain has come to the fore in our efforts to revisit biological phenomena from a fundamental perspective. Phenomenologically, this is attempted using the laws of classical physics, but during the past decade, several investigations pointed towards the possibility of harnessing quantum effects at biological scales as well. Biological systems, although, are nowhere close to the controlled environments created by ultra-cold atoms in ultrahigh vacuum, which allow manifestations of quantum coherence even at large timescales. Nevertheless, they are also not like the classical systems where the off-diagonal terms of the density matrix have been completely washed away. Indeed, several experiments of the last decade indicate that the quantum phenomena have played an important role in the progress of life~\cite{ball2011physics}. It is now argued that quantum effects may be intimately connected with the phenomena of Highly efficient Enzyme Catalysis and in the Development of Intelligence. For example, the high efficiency of Enzyme Catalysis is associated with the phenomena of quantum tunneling.

\subsection{A Few Final Thoughts on Emergence in Life Sciences}

Even with many such developments, we still face difficulties even in defining the emergent phenomena in the context of life sciences. Consider a simple example of a bacterial flagellar motor. Bacteria with flagella use them to swim, often rotating the flagellum about its base clockwise or anti-clockwise. The flagellum itself is attached to the motor, which is embedded into the bacterium's cell wall. The bacterial flagellum is itself just a part of a larger cell that, while it lives, consumes food, interacts with its environment, and must reproduce by simple fission. Thus, it satisfies all the criteria that we usually associate with living creatures many times their size. Additionally, there are also examples at the molecular scale of unusual organized behavior. A simple example is that of a molecular motor. A molecular motor is a protein molecule that breaks down another energy-carrying molecule, ATP. While doing this, it must work against the drag forces from the surrounding fluid, thus transducing the energy from the hydrolysis of ATP into work. Such motors, among them both translational and rotary motors, are highly efficient. In a secluded picture, all this is happening without any intervention, but if looked at from a holistic perspective, there is a well-defined purpose of sustaining the life, apportioning them the tag of emergent phenomena themselves.

Further, in the previous section, we have seen that the origin of life and the appearance of intelligence in living entities can be looked at as an emergent phenomenon.
Although we know that biology originates from the underlying physics, but we are still oblivious to any mechanism where causal connections can be established between the fundamental physical laws and the dynamics of life that we observe. Let us dwell on it: The theories of physics are devoid of any notion of purpose or function; for example, classical physics exhibits Hamiltonian dynamics where the evolution is deterministic and happens in a clockwork manner. While at the level of the living entities, their functioning is always for accomplishing some purpose. This indeed creates a big problem of reconciliation. Here, it is also essential to mention that although quantum physics defies the deterministic outcomes and exhibits a branching behavior (due to irreducible randomness), the outcomes are just a matter of chance, and no underlying mechanism of choice seems to be at work.  

Indeed, fathoming the emergence in biological contexts is unimaginably complex, and at times the pursuit appears to be generating more questions than answers, if only because there have been many choices to be made along the route to Homo sapiens. But perhaps the development of a holistic understanding of life forms may itself be an emergent reality in evolutionary timescale!

\section{Conclusion}

This article was an attempt to introduce the notions of collective behavior and emergent phenomena to the young researchers so that a holistic understanding of the macroscopic world around us can be promulgated. While focusing primarily on the instances of emergent behavior in condensed matter systems, we have also provided glimpses of the emergence across the various other domains of knowledge. This was attempted in order to broaden our perspectives and to appreciate the common interests of researchers belonging to seemingly unassociated disciplines, as we believe that convergence of ideas will be a key enabler for the development of science and technology in the coming times.

\section{Author Information}

Nitish Kumar Gupta (email-nitishkg@iitk.ac.in) is with Centre for Lasers \& Photonics, Indian Institute of Technology Kanpur, 208016, India.  A. M. Jayannavar (email-jayan@iopb.res.in) is with Institute of Physics, Bhubaneswar, Odisha 751005, India.

\section{Acknowledgement}

A. M. Jayannavar acknowledges DST, India for J C Bose fellowship.

\bibliographystyle{unsrt}
\bibliography{sample}

\end{document}